\documentclass[conference]{IEEEtran}
\IEEEoverridecommandlockouts

\usepackage{cite}
\usepackage{amsmath,amssymb,amsfonts}
\usepackage{algorithmic}
\usepackage{graphicx}
\usepackage{textcomp}
\usepackage{xcolor}
\usepackage{hyperref}
\usepackage{array,booktabs,longtable,tabularx}
\usepackage{ltablex}
\usepackage{siunitx}
\usepackage{tabulary}
\usepackage{float}
\restylefloat{table}
\usepackage{multirow} 
\usepackage{multicol}
\usepackage{cleveref}
\usepackage{amssymb}
\usepackage{verbatim}
\usepackage{amsmath}
\usepackage{subcaption}

\def\BibTeX{{\rm B\kern-.05em{\sc i\kern-.025em b}\kern-.08em
    T\kern-.1667em\lower.7ex\hbox{E}\kern-.125emX}}

\begin{document}

\title{Traffic Priority-Aware 5G NR-U/Wi-Fi Coexistence with Deep Reinforcement Learning \\
\thanks{This work was supported in part by the U.S. National Science Foundation under Grant 2034616.}
}

\author{
    \IEEEauthorblockN{Mohammad Reza Fasihi and
        Brian L. Mark}
    \IEEEauthorblockA{Dept. of Electrical and Computer Engineering
        and Wireless Cyber Center}
    \IEEEauthorblockA{George Mason University, Fairfax, Virginia, United States\\
        Email: mfasihi4@gmu.edu, bmark@gmu.edu}
}

\date{February 2024}

\maketitle

\begin{abstract}

\par Coexistence of 5G new radio unlicensed (NR-U) and Wi-Fi is highly prone to the collisions among NR-U gNBs (5G base stations) and Wi-Fi APs (access points). To improve performance and fairness for both networks, various collision resolution mechanisms have been proposed to replace the simple listen-before-talk (LBT) scheme used in the current 5G standard. We address two gaps in the literature: first, the lack of a comprehensive performance comparison among the proposed collision resolution mechanisms and second, the impact of multiple traffic priority classes. Through extensive simulations, we compare the performance of several recently proposed collision resolution mechanisms for NR-U/Wi-Fi coexistence. We extend one of these mechanisms to handle multiple traffic priorities. We then develop a traffic-aware multi-objective deep reinforcement learning algorithm for the scenario of coexistence of high-priority traffic gNB user equipment (UE) with multiple lower-priority traffic UEs and Wi-Fi stations. The objective is to ensure low latency for high-priority gNB traffic while increasing the airtime fairness among the NR-U and Wi-Fi networks. Our simulation results show that the proposed algorithm lowers the channel access delay of high-priority traffic while improving the fairness among both networks.
\end{abstract}


\IEEEpeerreviewmaketitle

\section{Introduction}
\label{sec:introduction}


\par Over the past decade, researchers have been actively exploring ways to ensure efficient coexistence between different radio access technologies (RATs) in the unlicensed spectrum\cite{Sathya:2021, Saha:2021, Hirzallah:2021,  Naik:2020, Sathya:2020}.  Wi-Fi technology, the main incumbent in this spectrum, employs Enhanced Distributed Channel Access (EDCA), which is based on Carrier Sense Multiple Access with Collision Avoidance (CSMA/CA). 3GPP release 16 introduced 5G New Radio Unlicensed (NR-U) as an improvement of LTE-LAA that allows 5G to operate in unlicensed spectrum. NR-U enables its base stations (gNBs) to operate in unlicensed bands other than 5~GHz, such as 3.5~GHz, 6~GHz, and 60~GHz. Like its predecessor, NR-U adheres to ETSI's listen-before-talk (LBT) channel access scheme for coexistence with Wi-Fi.

\par LBT is a simple method of resolving contention with other transmitting devices. However, the ending point of the LBT procedure may not coincide with NR-U's slot boundary and the 3GPP specification does not regulate the behavior of the gNB to prevent the channel from being occupied by other devices. One solution is for the gNB to block the channel with a reservation signal (RS) for the rest of the slot duration. Although this improves the performance of the 5G NR-U network, it degrades the Wi-Fi network's performance. Furthermore, the RS approach has been criticized as a potential source of inefficiency~\cite{Zajac:2022}. It is also possible that multiple transmitters finish their LBT procedures and start their transmissions at the same time, which will result in more collisions. More collisions in turn result in higher channel access delay for the transmitters, especially in a dense network scenario. Therefore, it is crucial to deploy an efficient \textit{collision resolution} mechanism. 

\par In Wi-Fi networks, stations (STAs) may use the Request to Send/Clear to Send (RTS/CTS) mechanism to protect their transmissions from collisions. However, NR-U lacks such a mechanism. The problem becomes worse when gNBs aggregate multiple channels for transmission and the RF power leakage from one channel to the adjacent channels causes the gNB to sense them busy even when
they are idle, thus deteriorating the aggregate capacity of the network. Although the RS significantly helps to solve this problem, it may lead to asymmetric collisions when Wi-Fi stations use RTS/CTS. Additionally, the RS does not carry any data and thus wastes channel resources. Therefore, alternative approaches have been proposed that aim to eliminate the drawbacks of the RS either by modifying the behavior of the gNB when using RS or by using a deterministic backoff (DB) mechanism~\cite{Zajac:2022, Loginov:2021, Loginov:2022, Szott:2022, Szott_2:2022, Kim:2021}. 

\par Three main service categories are defined in 5G new radio (NR): enhanced mobile broadband (eMBB), massive machine-type communication (mMTC), and ultra-reliable low-latency communication (URLLC). Among these service types, URLLC is the most challenging because it has two conflicting requirements: 3GPP Release 16 and 17 specify the
reliability up to $99.99999$ percent and the latency down to $0.5$~ms for URLLC~\cite{Le:2021}. The probability of inter/intra-network collisions for both Wi-Fi and NR-U networks may hinder their expected overall performance. Specifically, when transmitters with different traffic priorities contend for the channel access, the lower priority traffic may experience fewer collisions than higher priority traffic~\cite{Muhammad:2020}. None of the
existing collision resolution mechanisms proposed in the literature have considered the impact of multiple traffic priorities. 
In this paper, we develop a {\em traffic priority-aware} channel access mechanism for the coexistence of NR-U/Wi-Fi that can provide guaranteed delay performance for high-priority traffic while maximizing airtime fairness between two networks. 


\par The remainder of the paper is organized as follows. In Section~\ref{sec:comparison}, we compare the performance of the major collision resolution techniques proposed in the recent literature through extensive simulations. 
In Section~\ref{sec:traffic_aware}, we study the performance of the NR-U network when transmitters with different traffic classes share the unlicensed spectrum with the Wi-Fi network. Then, we propose two 
new traffic priority-aware collision algorithms based on 
extending the gCR-LBT protocol, which was found to have the best overall performance in Section~\ref{sec:comparison}.  The first uses a dynamic
transmission skipping method, which is effective in an NR-U only network.
The second employs multi-objective deep reinforcement learning 
to address the NR-U/Wi-Fi coexistence scenario.  In Section~\ref{sec:simulation_results}, we present simulation results, which
demonstrate that our proposed priority-aware
algorithms are able to protect the delay performance of high-priority traffic nodes when there are multiple lower-priority traffic nodes contending to access the channel.  The paper is concluded in Section~\ref{sec:conclusion}.


\section{Evaluation of Collision Resolution Schemes}
\label{sec:comparison}

\par In this section, we compare the performance of several collision resolution techniques proposed in the literature along with the legacy RS-LBT through 
extensive simulations. We extended the simulation model developed in~\cite{Zajac:2022} and~\cite{Jan:2022} by implementing RS-LBT, GAP-LBT~\cite{Zajac:2022}, CR-LBT~\cite{Loginov:2022}, gCR-LBT~\cite{Loginov:2021}, and DB-LBT~\cite{Szott:2022}, and integrating the Wi-Fi network to enable a realistic analysis of a coexistence scenario in shared sub-7 GHz spectrum. For each of the proposed multiple access techniques, we use the channel access parameter values from the associated reference. All transmitters are assumed to be full-buffered and the physical layer is assumed to be error-free. All transmitters are located in the transmission range of each other and there are no hidden terminals. The number of nodes for both NR-U and Wi-Fi networks are varied from 1 to 15. Channel access parameters of both networks follow the Best Effort Access Category except the transmission duration, which is considered to be $2$~ms for both Wi-Fi and NR-U to have a fair comparison. 

Simulation parameters are listed in Table~\ref{simulation_param}. 
The following metrics are used in the performance evaluation study:
\begin{itemize}
    \item \textbf{intra-network collision probability:} the ratio of failed transmissions to the total number of transmissions due to collisions between nodes of the same network;
    \item \textbf{channel efficiency:} the ratio of the total successful airtime of nodes of a network to the total simulation time;
    \item \textbf{channel access delay:} the average time between two consecutive successful transmissions over all nodes.
    \item \textbf{Jain's fairness index (JFI):} quantifies how fairly the channel airtime is divided among NR-U and Wi-Fi.
\end{itemize}

\begin{table}
    \centering
    \caption{Default simulation parameters.}
    \begin{tabular}{| c | c |}
        \hline
        \textbf{Parameter} & \textbf{Value} \\ 
        \hline
        Wi-Fi AP $\mbox{CW}_{\min}$, $\mbox{CW}_{\max}$ & $15, 63$ \\ 
        NR-U gNB $\mbox{CW}_{\min}$, $\mbox{CW}_{\max}$ & $15, 63$ \\ 
        Wi-Fi transmission duration & $2$~ms \\
        NR-U transmission duration & $2$~ms \\
        Synchronization slot duration & $500~\mu$s \\
        Simulation time & $10$~s \\
        \hline
    \end{tabular}
    \label{simulation_param}
\end{table}

\begin{figure*}
\begin{multicols}{2}
    \includegraphics[width=\linewidth]{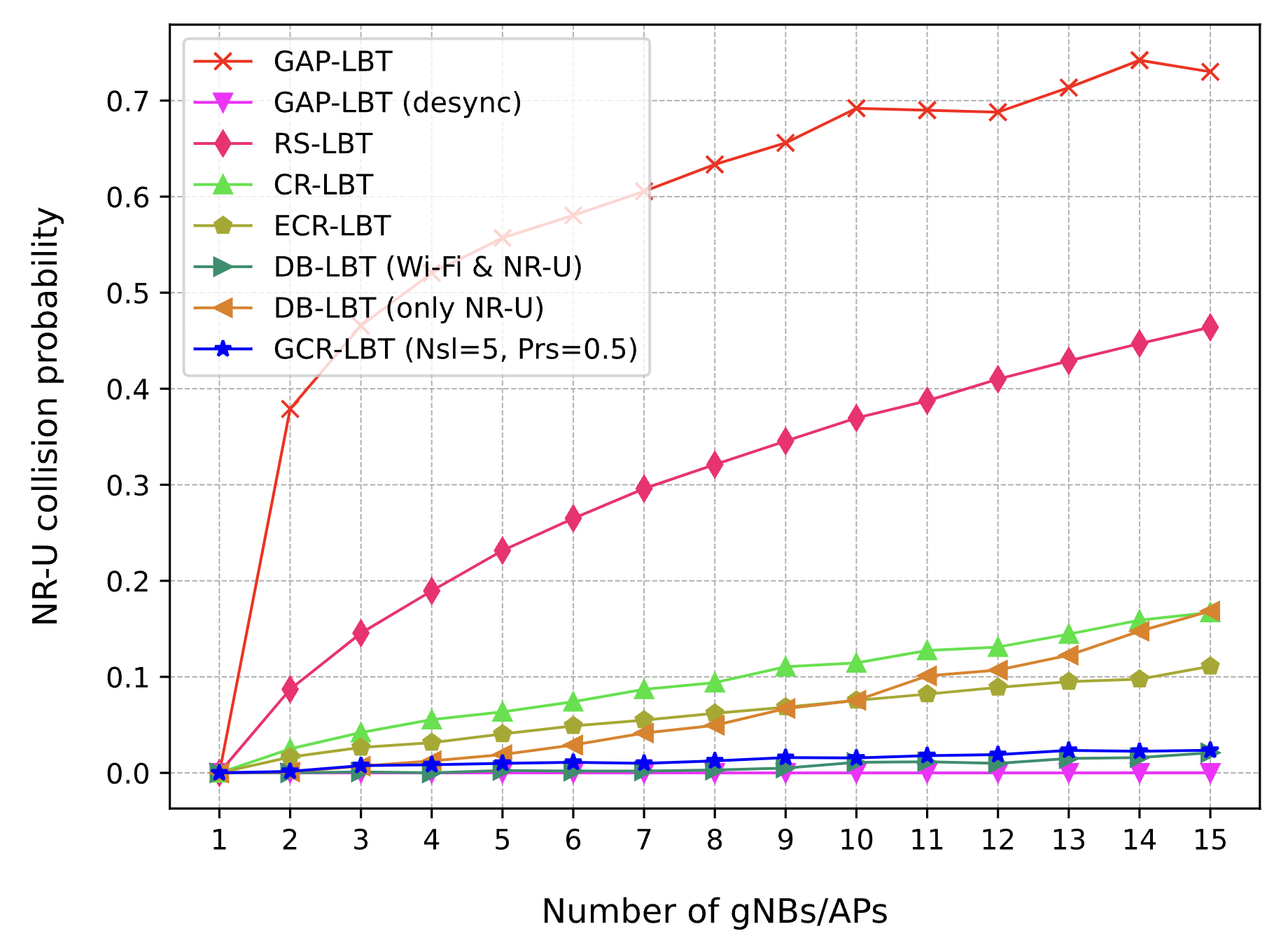}
    \caption{Intra-gNBs collisions probability for NR-U.}
    \label{nru_collision}
    \par
    \centering
    \includegraphics[width=\linewidth]{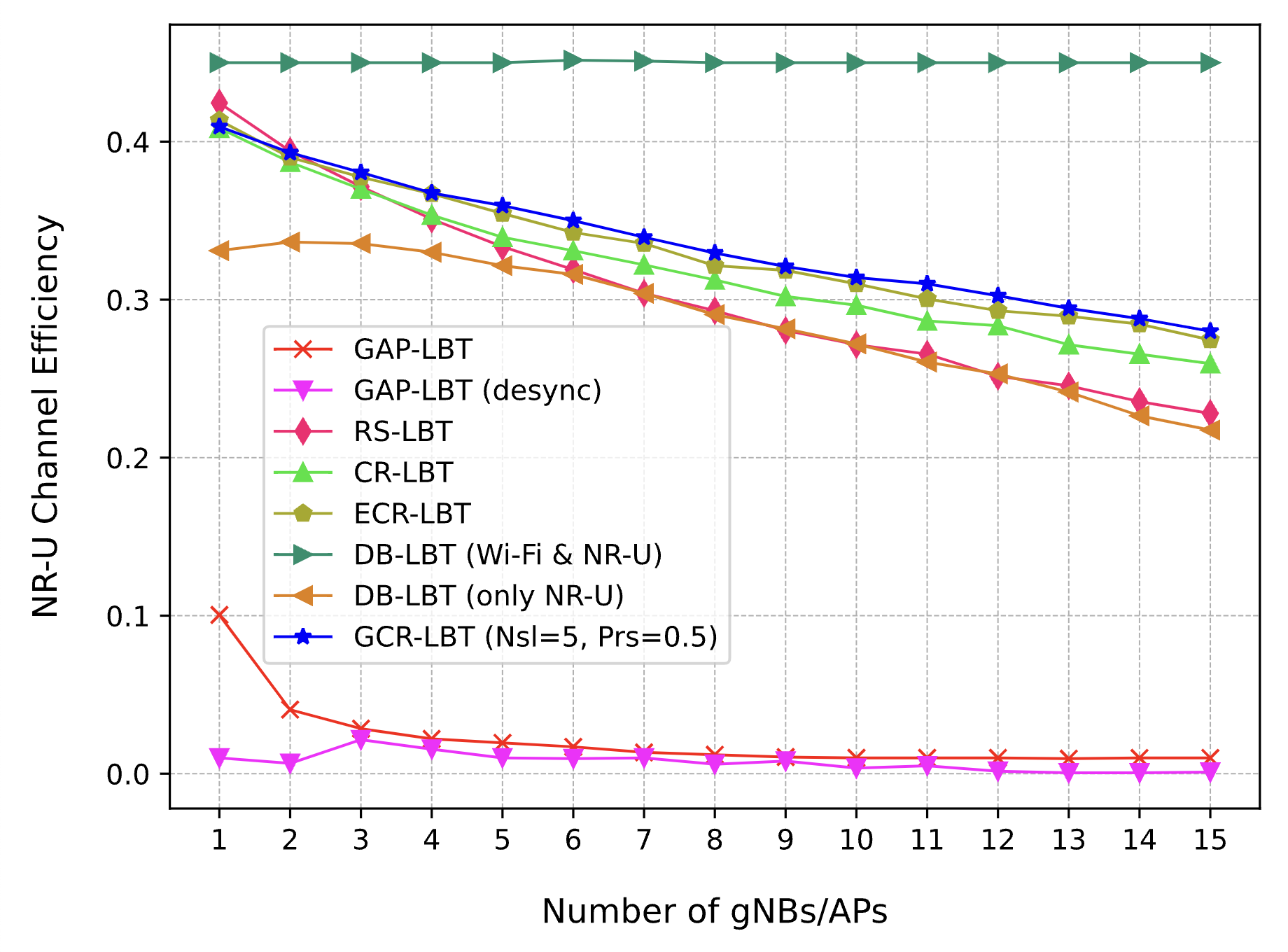}
    \caption{Channel efficiency of NR-U.}
    \label{nru_efficiency}
    \par
\end{multicols}
\begin{multicols}{2}
    \centering
    \includegraphics[width=\linewidth]{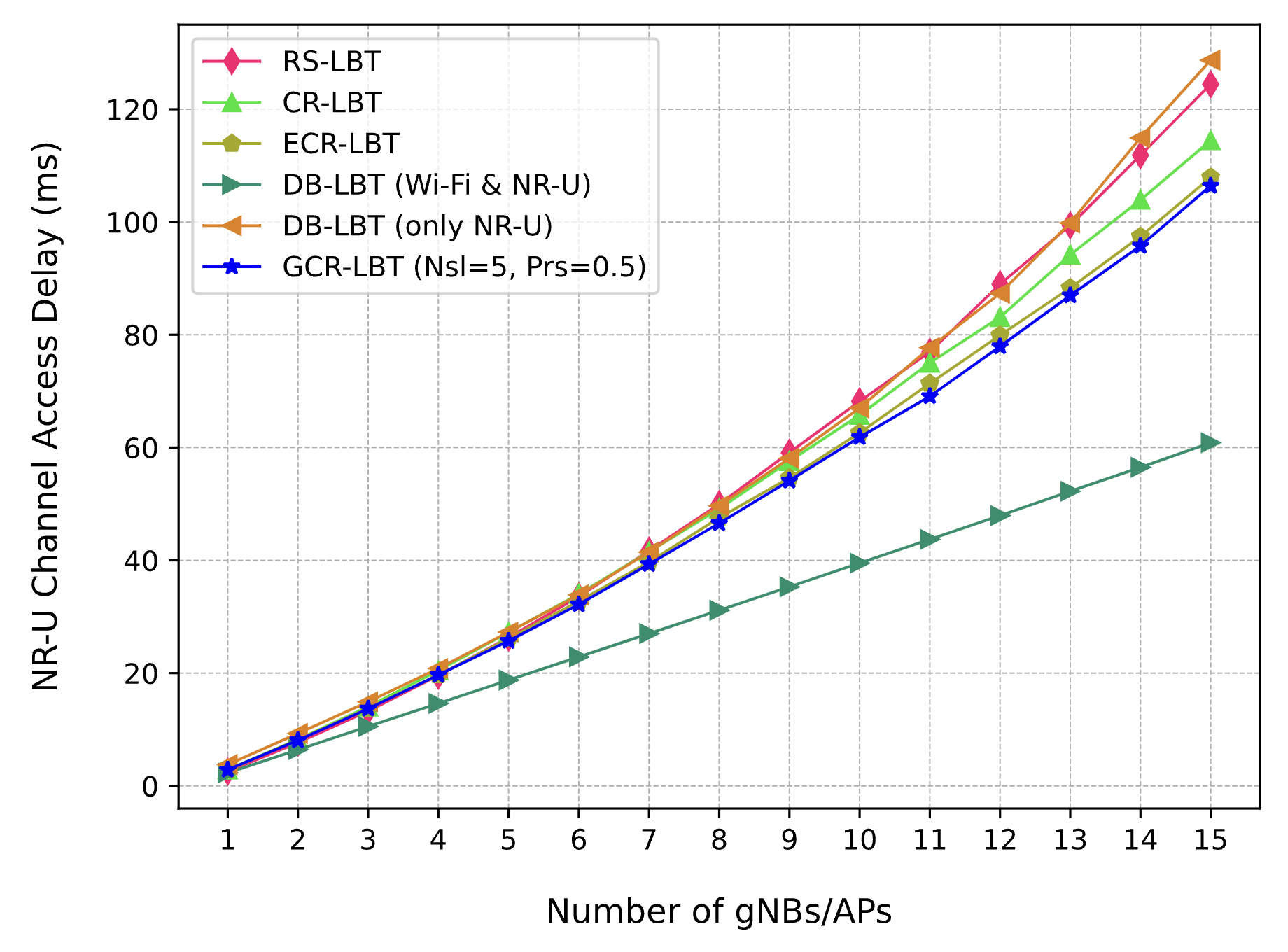}
    \caption{Channel access delay of NR-U.}
    \label{nru_delay}
    \par
    \includegraphics[width=\linewidth]{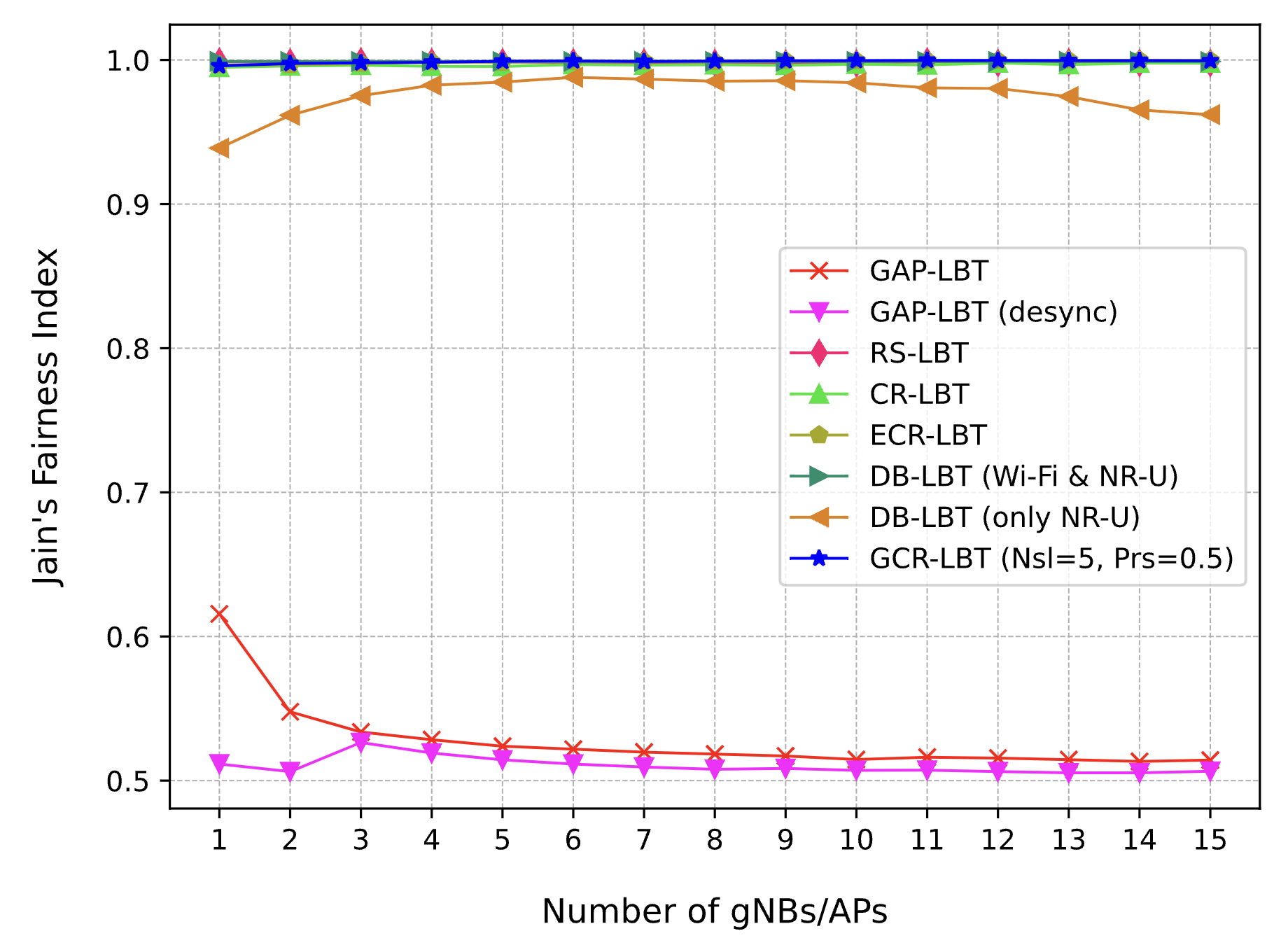}
    \caption{Jain's fairness index between Wi-Fi and NR-U.}
    \label{total_jfi}
    \par    
\end{multicols}
\end{figure*}

\par The intra-network collision probability of NR-U network for various collision resolution schemes is depicted in Fig.~\ref{nru_collision}. The GAP-LBT approach has the highest probability of collision when both networks are operating due to the restriction of beginning the transmission at licensed spectrum slot boundaries (LSSBs). One way to alleviate this is to desynchronize the slot boundaries of gNBs such that their slot boundaries do not coincide with each other~\cite{Zajac:2022}. The legacy RS-LBT has a lower collision probability than GAP-LBT, but it is still high compared to the other techniques due to the lack of a collision resolution mechanism for nodes that finish their backoff processes at the same time, which occurs more often when the number of nodes increases~\cite{Kim:2021}. The CR-LBT techniques (CR-LBT~\cite{Loginov:2022}, eCR-LBT and gCR-LBT~\cite{Loginov:2021}) modify the transmission of RS by introducing the concept of collision resolution slot (CR-slot), which significantly reduces the probability of collision among gNBs. As  shown in Fig.~\ref{nru_collision}, the CR-LBT techniques show good collision resolution performance. Among them, gCR-LBT has slightly better performance with respect to the mini-slot transmission of NR-U. DB-LBT extends the idea of deterministic backoff (DB) for Wi-Fi networks~\cite{Wentink:2017} to NR-U networks. It schedules all the transmitters in a round-robin fashion and keeps the scheduling fixed until the set of active transmitters changes. The initial backoff value for all transmitters is adjusted to allow the new transmitters to be scheduled for channel access~\cite{Szott:2022}.

\par Fig.~\ref{nru_efficiency} shows the channel efficiency of 
the considered channel access schemes. GAP-LBT has the worst performance. Although desynchronization significantly reduces the number of collisions, the lack of a channel reservation mechanism severely degrades the efficiency of GAP-LBT in coexistence scenarios.  The Wi-Fi APs start their transmissions immediately after finishing the backoff process, whereas gNBs have to wait for the LSSB. RS-LBT has better channel efficiency performance than GAP-LBT, but it is still worse than that of the other schemes, due to the lack of
a collision resolution scheme. Among the family of CR-LBT schemes, gCR-LBT has better collision resolution performance than the others. When {\em both} networks employ the DB method, DB-LBT has the best channel efficiency performance as it schedules all nodes in a round-robin fashion. Hence, when the backoff values of all nodes converge to the same value after several contention rounds, there is no wastage of channel airtime. 

Fig.~\ref{nru_delay} shows the channel access delay of different techniques. We exclude GAP-LBT from this figure as it has much higher delay than the others. RS-LBT has the highest delay among the remaining methods due to the high probability of collision. DB-LBT, when only the NR-U network performs the DB, also has a high channel access delay. Again, gCR-LBT has the best delay performance among other techniques which makes it a good candidate for delay-sensitive applications when we can only modify the channel access mechanism of the NR-U network. The channel access delay of all methods except DB-LBT increases exponentially as the number of nodes increases, but for DB-LBT with both networks performing DB, the delay increases linearly with the number of active nodes. 
Fig.~\ref{total_jfi} shows the JFI for channel airtime of Wi-Fi and NR-U networks. It is obvious that all methods maintain very good fairness between two networks except GAP-LBT, which lacks channel reservation and collision resolution mechanisms. 

\par \textit{We conclude that gCR-LBT has the best overall performance when only the channel access mechanism of the NR-U network can be specified by the operator. Thus, we adopt gCR-LBT as the underlying collision resolution protocol for our priority-aware channel access schemes developed in the next section.}

\begin{figure*}
  \includegraphics[width=\linewidth]{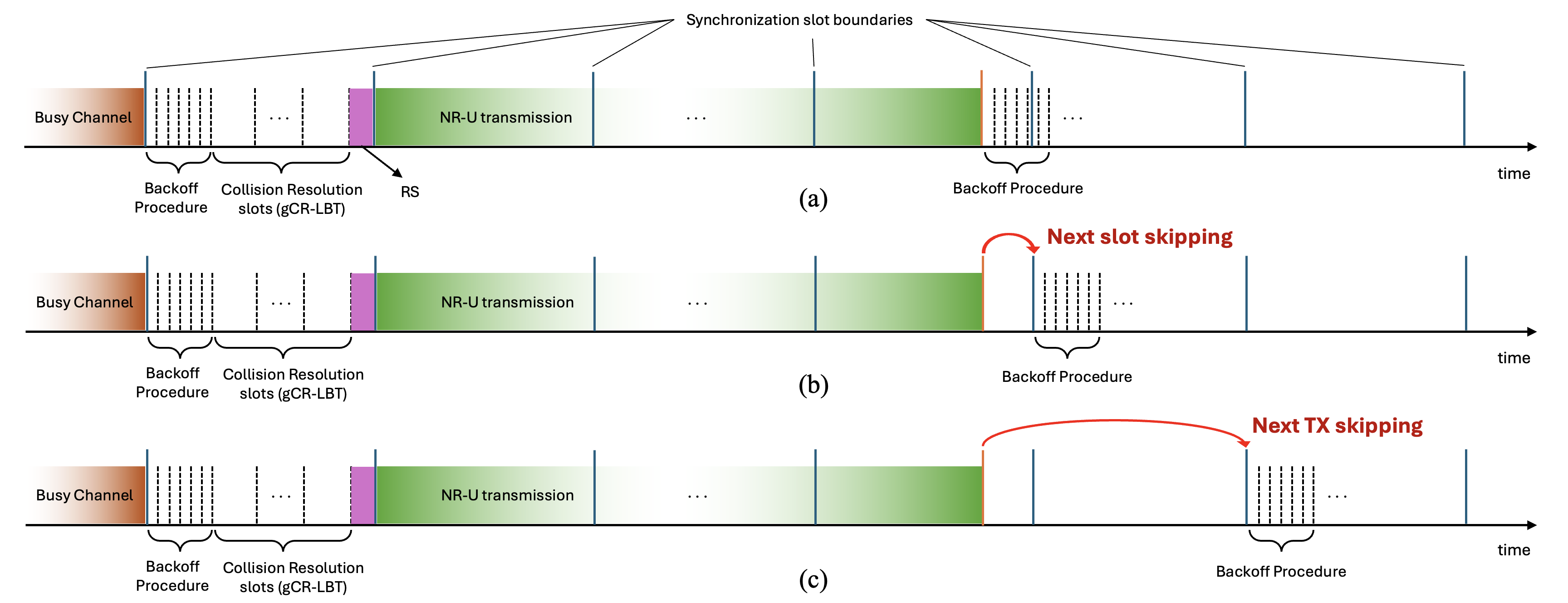}
  \caption{Example of gNB UE running gCR-LBT~\cite{Loginov:2021} with a) no skipping, b) skipping to the next slot boundary, and c) skipping to the next transmission opportunity.}
  \label{tx_skipping}
\end{figure*}

\section{Traffic Priority-Aware Channel Access}
\label{sec:traffic_aware}

\par A transmitter must perform the LBT procedure to access a channel in an unlicensed spectrum. This can harm the latency performance of high-priority traffic such as URLLC packets, especially when contending for channel access with low-priority traffic. In addition, the LBT method may cause additional channel access delay compared to the licensed spectrum due to the unpredictability of the transmission opportunity. Therefore, the delay performance of high-priority traffic may not meet URLLC requirements. In this section, we propose methods to protect the low-latency performance of high-priority transmitters while maintaining high fairness in two scenarios: NR-U only and NR-U/Wi-Fi coexistence. We denote the high-priority and low-priority transmitters by PC1 (priority class 1) and PC3 (priority class 3), respectively.

\subsection{Dynamic Transmission Skipping Method}

We propose a \textit{Dynamic Transmission Skipping} method to handle PC1 and PC3 traffic. In this method, the gNB asks the PC3 transmitter that has just finished its successful transmission to defer its next transmission attempt to the next slot boundary or to skip its future transmission opportunities in order to allow PC1 transmitters to have faster channel access. To be specific, at the beginning of each radio frame, the gNB observes the channel access delay of PC1 transmitters and compares it with the URLLC delay requirement. If the delay is more than what is required, the gNB forces the PC3 transmitter to skip its next transmission attempts to the next licensed spectrum slot boundary (LSSB) after finishing a successful transmission. If the delay requirement is still not met, it forces the PC3 transmitter to skip its next opportunity to transmit (i.e., the next viable slot in which it could attempt a transmission)~\cite{Zajac:2022}. The gNB repeats this behavior at the beginning of each frame until the URLLC requirement is met. Fig.~\ref{tx_skipping} shows the concept of transmission skipping. To maintain a high degree of fairness, we assume that all gNBs employ gCR-LBT (see Section~\ref{sec:comparison}). 

\begin{figure*}[t]
\begin{multicols}{2}
\centering
    \includegraphics[width=\linewidth]{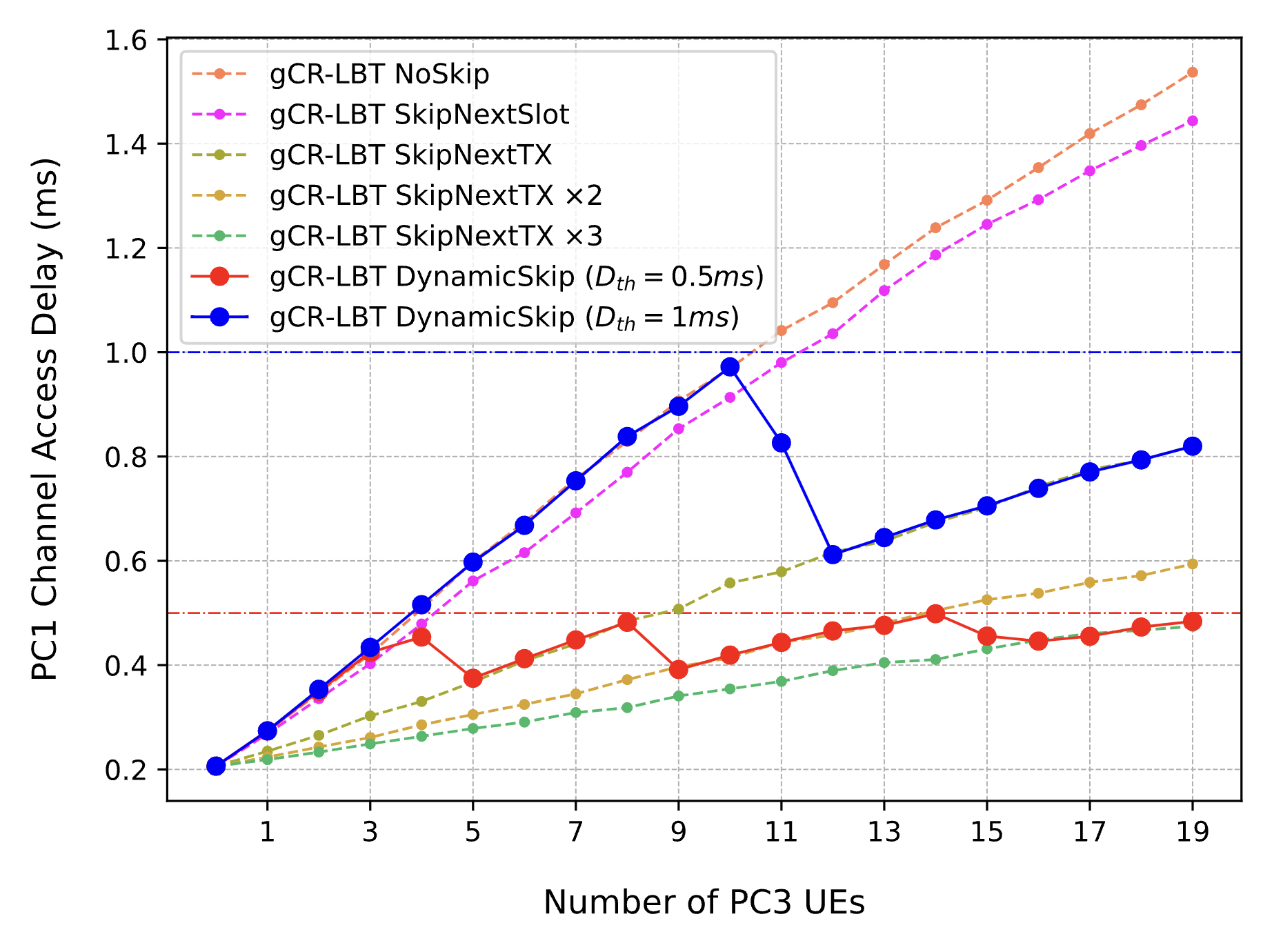}
        \caption{Channel access delay of PC1 transmitter (NR-U only scenario) with different number of transmission skips and dynamic skipping for PC3 transmitters.}
        \label{dynamic_skip_delay_pc1}
        \par
    \includegraphics[width=\linewidth]{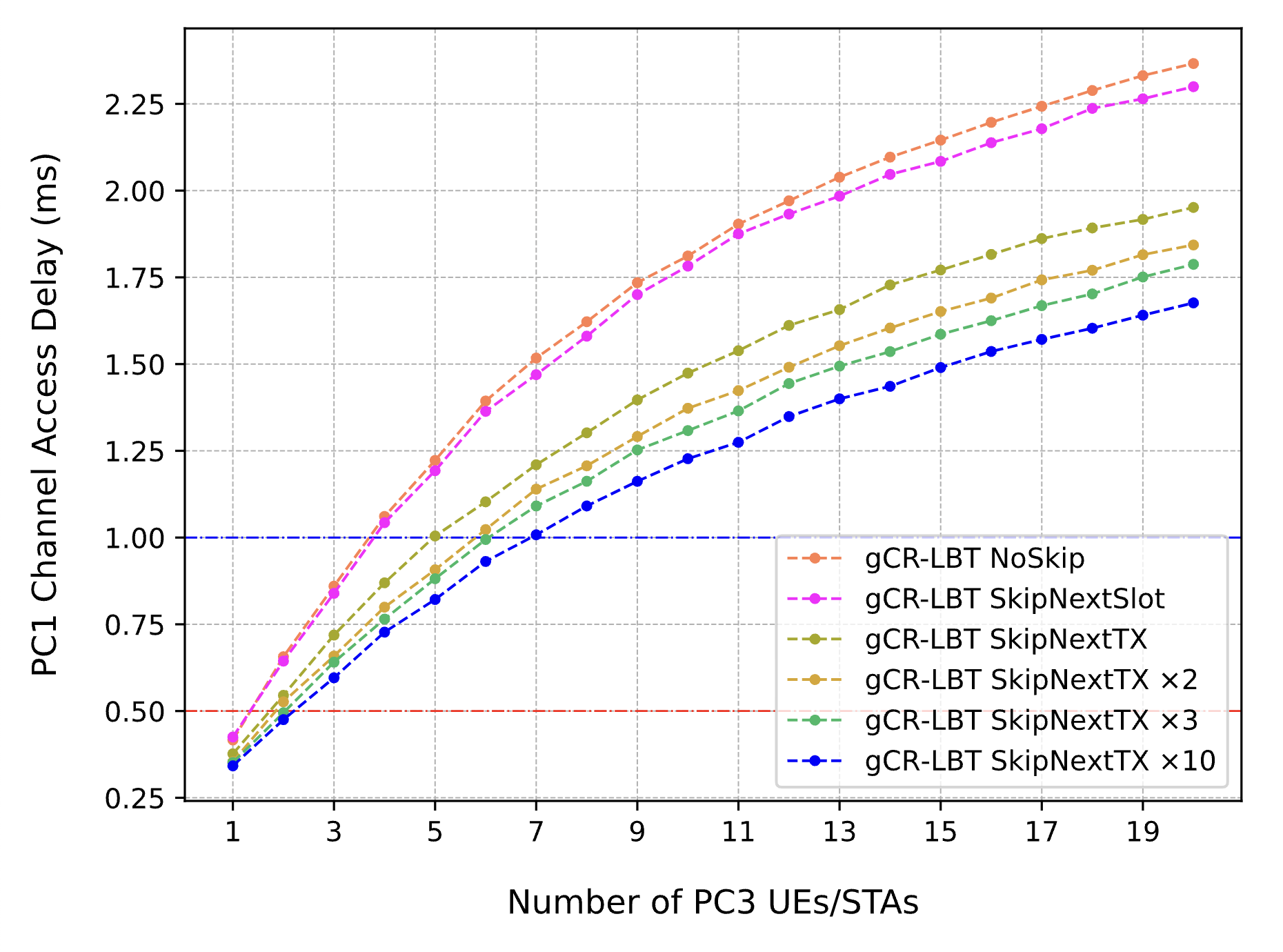}
        \caption{Channel access delay of PC1 transmitter (Wi-Fi/NR-U coexistence scenario) with different number of transmission skips for PC3 transmitters.}
        \label{pc1_delay_coexistence}
        \par
\end{multicols}
\end{figure*}

\par Fig.~\ref{dynamic_skip_delay_pc1} shows the channel access delay of one PC1 transmitter when it shares the channel with varying numbers of PC3 transmitters. The dashed lines represent the channel access delay of PC1 node when PC3 nodes do no skip (\textit{NoSkip}) or consistently choose to skip their next transmission attempts until the next slot boundary (\textit{SkipNextSlot}), the next transmission opportunity (\textit{SkipNextTX}), next two transmission opportunities (\textit{SkipNextTX$\times2$}), and next three transmission opportunities (\textit{SkipNextTX$\times3$}) after a successful transmission. 
By skipping transmission opportunities, the PC3 transmitters allow
more opportunities for the PC1 node, resulting in lower channel access
delay for the PC1 node. On the other hand, as is shown in the figure, the amount of PC1 delay increases linearly with the number of PC3 nodes. Thus, the delay may exceed the maximum allowed channel access delay for URLLC after the total number of PC3 nodes increases beyond a threshold. To avoid this, at the beginning of each radio frame, the gNB initiates dynamic transmission skipping by 
informing the PC3 nodes how many transmission opportunities to skip in order to protect the URLLC transmission. The performance of the dynamic skipping method is shown by solid lines for maximum allowed delays of $0.5$~ms and $1$~ms (shown by horizontal dashed lines) in Fig.~\ref{dynamic_skip_delay_pc1}.

\subsection{Multi-Objective DQN Method} 

\par In the NR-U/Wi-Fi coexistence scenario, due to the lack of control over the operation of Wi-Fi transmitters, the dynamic skipping method may not be beneficial. This is because Wi-Fi transmitters are more likely to have access to the channel since gNB transmitters skip their next transmission opportunities, and hence, the channel access delay of PC1 node may violate the URLLC requirement
even when the PC3 nodes skip a large number of their next transmission opportunities. This can be observed in Fig.~\ref{pc1_delay_coexistence}. Therefore, we propose a reinforcement learning (RL) approach in which the PC1 channel access delay and airtime efficiency information is exchanged between the NR-U and Wi-Fi networks.

\par The goal of the RL algorithm is to maintain the low-latency performance of the NR-U PC1 node while maximizing the overall fairness among NR-U UEs and Wi-Fi STAs. We again assume gCR-LBT as the underlying collision resolution method for NR-U. We propose a multi-objective deep Q-learning network (MO-DQN) algorithm~\cite{Moffaert:2013} to ensure that the delay of the PC1 node remains under a threshold when the number of contending PC3 UEs and STAs increases\footnote{Due to space limitations, we omit a discussion of the rationale for using the DQN in our RL approach.}. In our model, the gNB of the 5G NR-U network and AP of the Wi-Fi network are considered as the agents. The tuple $[\mathrm{E}_{w}, \mathrm{E}_{l,pc3}, \mathrm{D}_{l,pc1}, \mathrm{CW}_{w}, \mathrm{CW}_{l}]$ is selected as the state, which is the input to the DQN. The parameters
$\mathrm{E}_{w}$, $\mathrm{E}_{l,pc3}$, and $\mathrm{D}_{l,pc1}$ denote the Wi-Fi efficiency, efficiency of NR-U PC3 nodes, and percentage of NR-U PC1 node delay to the total delay of NR-U network, respectively.
The parameters $\mathrm{CW}_{w}$ and $\mathrm{CW}_{l}$ represent the contention window (CW) size of Wi-Fi and NR-U networks, respectively. We assume the efficiency, delay, and CW value information of both systems are accessible to both networks. The action of each agent corresponds directly to setting the new CW value and can be chosen from the discrete range $a\in \{ 0, 1, \ldots, 6 \}$. The output that is  used by each agent to update CW is then given by $\mbox{CW} = 2^{a+4}-1$, corresponding to a range
between 15 and 1023. 

\par To deal with multiple objectives in RL, we use the \textit{scalarization method} in which separate objective functions are first normalized to be in 
the range $[0, 1]$ and then aggregated into a single objective function by a non-decreasing aggregation function~\cite{Moffaert:2013}. Our goal is to decrease the PC1 channel access delay while increasing the JFI among both networks. Using the scalarization method, we define the aggregated objective function $f(a)$ and our multi-objective optimization problem as follows:
\begin{equation}
    \max_{a \in \{ 0, \ldots, 6 \}} f(a) = \alpha (1 - D_{l,pc1}(a)) + (1-\alpha) F_{\rm fair}(a) ,
    \label{reward_function}
\end{equation}
where $F_{\rm fair}(a)$ is the JFI among both networks and $\alpha \in [0.5, 1]$ is a tuning parameter that enables changing the weights associated with the channel access delay of PC1 and total JFI in
$f(a)$, respectively.

\begin{table}
    \centering
    \caption{Hyper-parameters of DQN}
    \begin{tabular}{| c | c |}
        \hline
        \textbf{Parameter} & \textbf{Value} \\ 
        \hline
        Interaction period & $10$~ms \\
        Discount factor & $0.7$ \\ 
        Range of $\epsilon$ & $0.9$ to $0.001$ \\ 
        DQN learning rate & $10^{-3}$ \\
        Batch size & 32 \\
        Dimensions of NN layers & $128 \times 64 \times 7$\\
        \hline
    \end{tabular}
    \label{DQN_param}
\end{table}

\begin{figure*}
\begin{multicols}{2}
\centering
    \includegraphics[width=\linewidth]{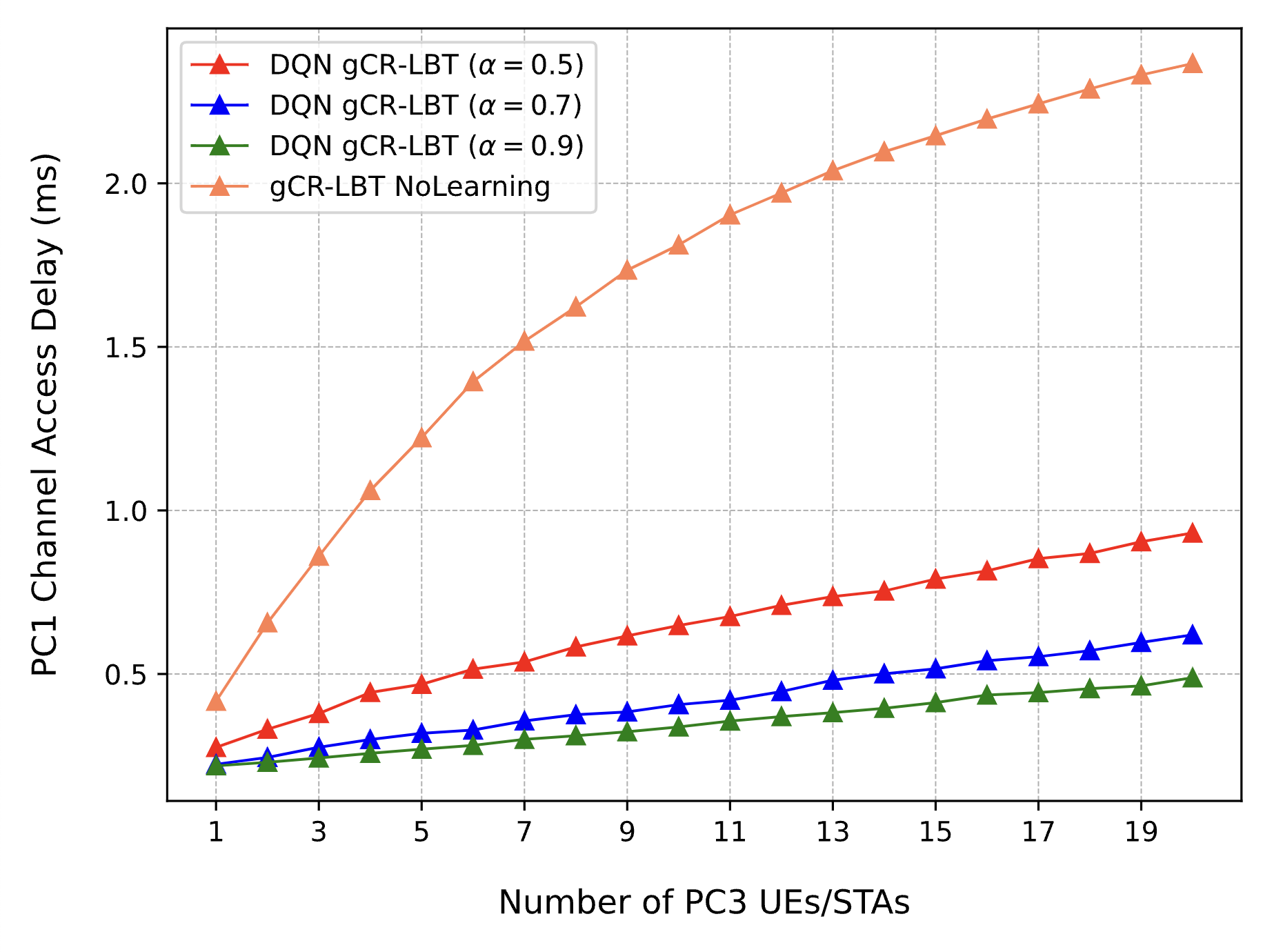}
        \caption{Channel access delay of high-priority (PC1) UE.}
        \label{dqn_delay}
    \par
	\includegraphics[width=\linewidth]{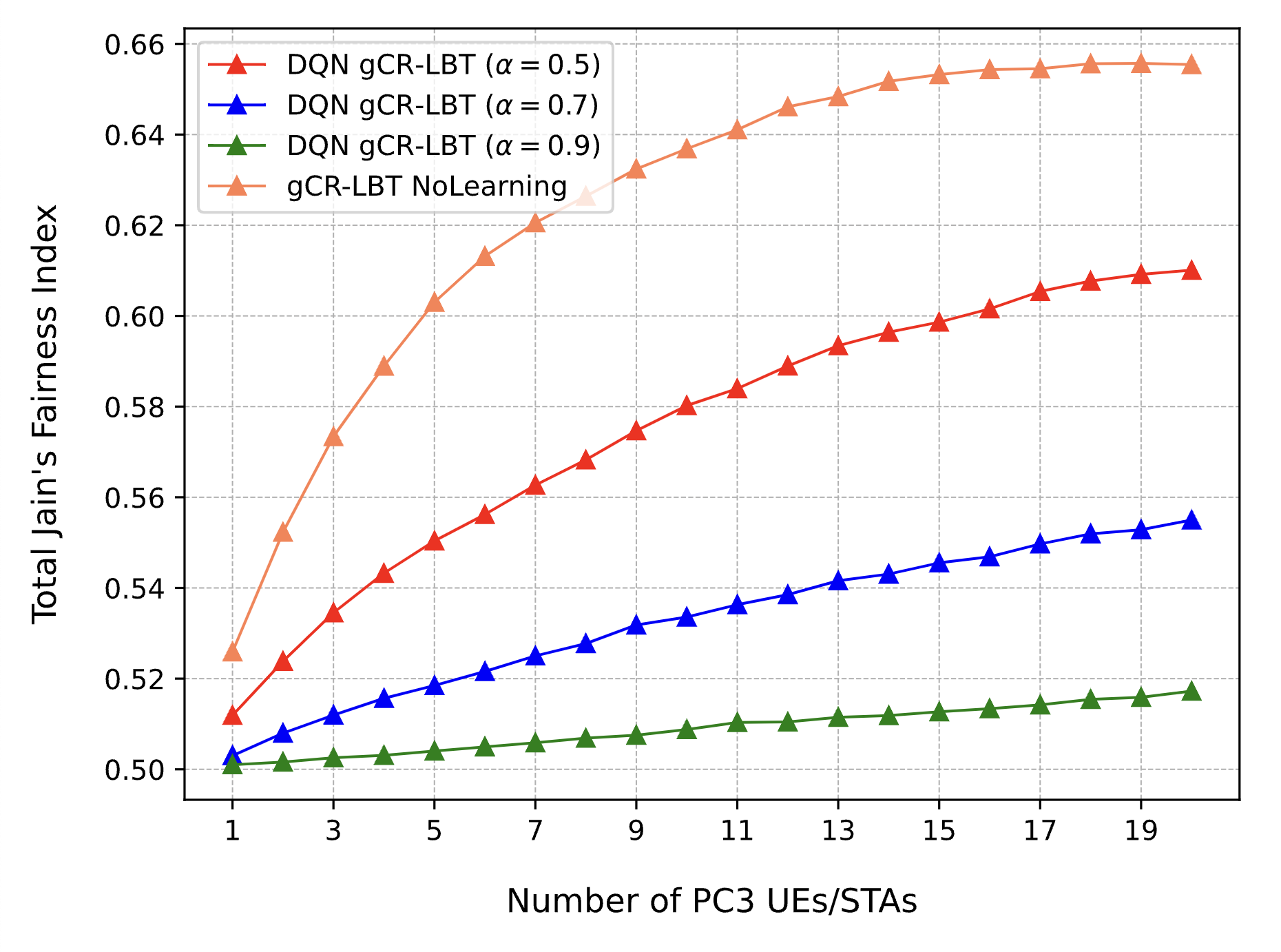}
    	\caption{JFI among Wi-Fi and NR-U.}
    	\label{dqn_jfi}
    \par
\end{multicols}
\end{figure*}

\par In the training phase of the deep Q-learning network (DQN), we generate random weights for the Q-value estimator neural network (NN). Each agent makes an observation and chooses its action based on the $\epsilon$-greedy method wherein the exploration rate decays from 0.9 to 0.001 in $20\%$ of the total iterations to enable a smooth transition from pre-learning phase to learning phase. After collecting enough training data, we adopt the \textit{Adam} (Adaptive Moment Estimation) optimizer~\cite{Kingma:2015} for training the Q-value estimator NN. Following the learning phase, the trained NN with weight vector $\boldsymbol\theta$ is used for the prediction of Q-values at each state and actions are taken by agents according to 
\begin{align}
  a_{t} = \mbox{arg}~\max_{a^{\prime} \in \{0, \ldots, 6 \}}
  Q(s_t,a^{\prime};\boldsymbol{\theta}_t) ,  
\end{align}
where $Q(s_t,a; \boldsymbol{\theta}_t)$ denotes the state-action value function which is the expected discounted reward when starting in state $s$ and selecting an action $a$ at time $t$, and $\boldsymbol{\theta}_t$ is the weight vector of the Q-value estimator NN at time $t$. Our DQN structure consists of an input layer followed by three fully connected layers that output the predicted Q-value corresponding to the input action. The hyperparameters of our DQN algorithm are summarized in Table~\ref{DQN_param}.

\section{Simulation Results}
\label{sec:simulation_results}

\par Next, we highlight the performance of our proposed DQN-based, priority-aware channel access protocol through simulation results. For the simulation scenario, we consider one PC1 UE sharing the channel with equal numbers of PC3 UEs and Wi-Fi STAs (varying from 1 to 20), and compare the performance of the proposed algorithm with that of gCR-LBT with no learning for different values of $\alpha$ in (\ref{reward_function}). Fig.~\ref{dqn_delay} shows the channel access delay of PC1 UE for several values of $\alpha$. Our results show that increasing the value of $\alpha$ results in decreasing channel access delay for PC1 node. As $\alpha$ increases, the high-priority node has more access to the channel and hence has the lower channel access delay. Finally, the JFI among NR-U and Wi-Fi transmitters is shown in Fig.~\ref{dqn_jfi}. This shows that by increasing the value of $\alpha$, more weight is given to the the PC1 delay in (\ref{reward_function}) and hence, the network is less fair.

\section{Conclusion}
\label{sec:conclusion}

\par Through extensive simulations, we compared the performance of
several existing collision resolution proposed in the recent literature for 5G NR-U/Wi-Fi coexistence. Our results showed that gCR-LBT had the best overall performance among these techniques. We extended the gCR-LBT scheme to the scenario of coexistence of high-priority UE with multiple lower-priority traffic UEs and Wi-Fi STAs.  We then developed a traffic-aware multi-objective deep Q-learning algorithm to ensure low-latency performance for high-priority traffic while increasing the total airtime fairness. Our simulation results confirmed that the algorithm is able to maintain low channel access delay for high-priority traffic while maintaining high fairness. In ongoing work, we are extending our traffic-aware deep
reinforcement learning algorithm to the multi-channel scenario, which involves tuning physical layer parameters to combat the negative effects of channel leakage. 

\bibliographystyle{IEEEtran}
\bibliography{IEEEabrv,main}

\end{document}